\documentclass[english,pra,twocolumn,showpacs,amsmath,amssymb,lengthcheck]{revtex4-1}

\usepackage[normalem]{ulem}
\usepackage{hyperref}
\usepackage{graphicx}
\usepackage[]{color}
\usepackage{braket}

\def\be{\begin{equation}}
\def\ee{\end{equation}}
\def\bea{\begin{eqnarray}}
\def\eea{\end{eqnarray}}
\def\ba{\begin{array}}
\def\ea{\end{array}}

\def\om{\omega}

\def\nn{\nonumber}
\def\ket{\rangle}
\def\bra{\langle}

\def\fr{\frac}

\def\ra{\rightarrow}
\def\rm{\mathrm}

\newcommand{\mv}[1]{\langle #1\rangle}

\def\ua{\uparrow}
\def\da{\downarrow}

\begin{document}
\title{Ultracold fermions in a one-dimensional bipartite optical lattice: metal-insulator transitions driven by shaking}
\author{M. Di Liberto$^1$}
\author{D. Malpetti$^{1,2}$}
\author{G. I. Japaridze$^{3,4}$}
\author{C. Morais Smith$^1$}
\affiliation{$^1$Institute for Theoretical Physics, Center for Extreme Matter and Emergent Phenomena, Utrecht University, Leuvenlaan 4, 3584CE Utrecht, the Netherlands}
\affiliation{$^2$Department of Physics, University of Pavia-CNISM, Via Bassi 6, I-27100 Pavia, Italy}
\affiliation{$\mbox{}^3$ Andronikashvili Institute of Physics,
Tamarashvili 6, 0177 Tbilisi, Georgia}
\affiliation{$\mbox{}^4$ Ilia State University, Cholokasvili Avenue 3-5, 0162 Tbilisi, Georgia}
\date{\today}

\begin{abstract}
We theoretically investigate the behavior of a system of fermionic atoms loaded in a bipartite one-dimensional optical lattice that is under the action of an external time-periodic driving force. By using Floquet theory, an effective model is derived. The bare hopping coefficients are renormalized by zeroth order Bessel functions of the first kind with different arguments for the nearest-neighbor and next-nearest neighbor hopping. The insulating behavior characterizing the system at half-filling in the absence of driving is dynamically suppressed and for particular values of the driving parameter the system becomes either a standard metal or an unconventional metal with four Fermi points. The existence of the four Fermi-point metal relies on the fact that, as a consequence of the shaking procedure, the next-nearest-neighbor hopping coefficients become significant compared to the nearest-neighbor ones. We use the bosonization technique to investigate the effect of on-site Hubbard interactions on the four Fermi-point metal-insulator phase transition. Attractive interactions are expected to enlarge the regime of parameters where the unconventional metallic phase arises, whereas repulsive interactions reduce it. This metallic phase is known to be a Luther-Emery liquid (spin gapped metal) for both, repulsive and attractive interactions, contrarily to the usual Hubbard model which exhibits a Mott insulator phase  for repulsive interactions. Ultracold fermions in driven one-dimensional bipartite optical lattices provide an interesting platform for the realization of this long studied four Fermi-point unconventional metal. 
\end{abstract}


\maketitle


\section{Introduction}

In recent years, cold atoms in optical lattices have become a powerful tool for investigating quantum phase transitions and realizing new  and unconventional states of matter \cite{bloch1,lewenstein,bloch2}. Since the observation of the superfluid-Mott insulator (SF-MI) phase transition for the Bose-Hubbard model \cite{jaksch,greiner}, many models have been experimentally engineered and investigated with unprecedented control.

By introducing external time-dependent driving forces that dynamically suppress the hopping, namely by shaking the optical lattice \cite{lignier2007,goldman2014}, the SF-MI phase transition has been achieved without the need of controlling the lattice potential depth \cite{eckardt2005,creffield,zenesini2009}. Since then, the shaking technique has been employed in many other experimental setups to realize, for instance, classical magnetism \cite{struck2011}, artificial gauge potentials in one \cite{struck2012} and two \cite{struck2013} dimensions, extended ferromagnetic domains \cite{parker}, to control photon-assisted \cite{ma2011} and correlated tunneling \cite{teichmann2009,chen2011}, to generate super Bloch oscillations \cite{haller2010} and has inspired theoretical works that proposed schemes to realize doublon-holon condensates \cite{rapp2012}, non-Abelian gauge fields \cite{hauke2012}, density-dependent gauge potentials \cite{santos2013}, and correlated-hopping models \cite{diliberto2014}.

The high freedom available for generating optical lattices has also allowed one to play with the lattice geometry and to create bipartite lattices, which turned out to be a key ingredient to achieve higher-band condensates \cite{hemmerich2011,hemmerich2011PRL,olschlager}, coherence control \cite{dilibertocomparin2014}, density-wave dynamics \cite{trotzky2012}, graphene-like physics \cite{tarruell2012,uehlinger2013}, and to measure the Zak phase characterizing topological Bloch bands \cite{atala}.

In condensed-matter systems, the model of correlated
electrons in bipartite lattices with staggered on-site potential,
known as the ionic-Hubbard model, has been the subject of intensive
studies during the last decades
\cite{nag,Egami,res,fab,tor,brune,man,aligia2,Japaridze_MH}.
Initially, the ionic-Hubbard model was proposed to study organic
mixed-stack charge-transfer crystals \cite{nag} and later it has been
used to describe the ferroelectric transition in perovskite
materials \cite{Egami}. Intensive interest in the study of the
low-dimensional versions of the ionic-Hubbard model was motivated by
the extremely rich phase diagram of this model revealing, at
half-filling, the possibility for the realization of the band-insulator to
Mott-insulator quantum phase transition with increasing on-site
Hubbard coupling, via the sequence of unconventional insulating and/or metallic  phases
\cite{res,fab,tor,brune,man,aligia2,Japaridze_MH}.

A similar, but different mechanism for the realization
of the band-insulating state in the one-dimensional half-filled
electron system  has been proposed by Peierls in the early 30s of the last
century, via the alternation in magnitude of the nearest-neighbor hopping
amplitude \cite{Peierls}. However, contrary to the ionic-Hubbard
model, the behavior of the Peierls model smoothly depends
on the on-site Hubbard coupling and no quantum phase transitions are realized. 
Instead, one just finds a crossover from a band-insulating phase at weak coupling into
the spin-Peierls phase at strong repulsive interaction
\cite{Kivelson,Baerswyl,Campbell}. Therefore, less attention has been
given to the search of quantum phase transitions in the Peierls
insulator.

In this paper, we study a driven 1D
bipartite optical lattice half-filled with fermionic atoms and show that it
is possible to drive (band)-insulator to metal transitions by tuning
the shaking parameter. Due to the presence of the $A-B$ sublattice
characterized by nearest-neighbor hopping coefficients alternating
in magnitude, the half-filled system is a
Peierls insulator. Shaking the optical lattice at high-frequencies leads to a model
with effective hopping parameters, where the bare value is multiplied by a Bessel
function. Since the relevant hopping parameters are renormalized in
different ways, the system realizes a large variety of quantum
phases, such as several metals characterized by
a Fermi surface with four Fermi points or two Fermi points,
and Peierls insulators with direct or indirect gaps.

The paper is organized as follows. In Sec.~II, we introduce the bipartite
optical potential that we study and show the lowest two bands obtained by solving the corresponding Schr\"{o}dinger equation. In Sec.~III we derive a minimal tight-binding model describing the two lowest bands, discuss its symmetries and estimate its main parameters. In Sec.~IV, the Floquet theory is applied to the time-dependent problem of the driven optical potential and the effective Hamiltonian for the quasienergy spectrum is derived. In Sec.~V we discuss the quasienergy spectrum as a function of the shaking parameter and the main scenarios are presented. In Sec.~VI the half-filled phase diagram is analyzed. In Sec.~VII we comment on the effect of on-site interactions and in Sec.~VIII we draw our conclusions.


\section{Optical potential}

\begin{figure}[!htbp]
\centering
\includegraphics[width=0.45\textwidth]{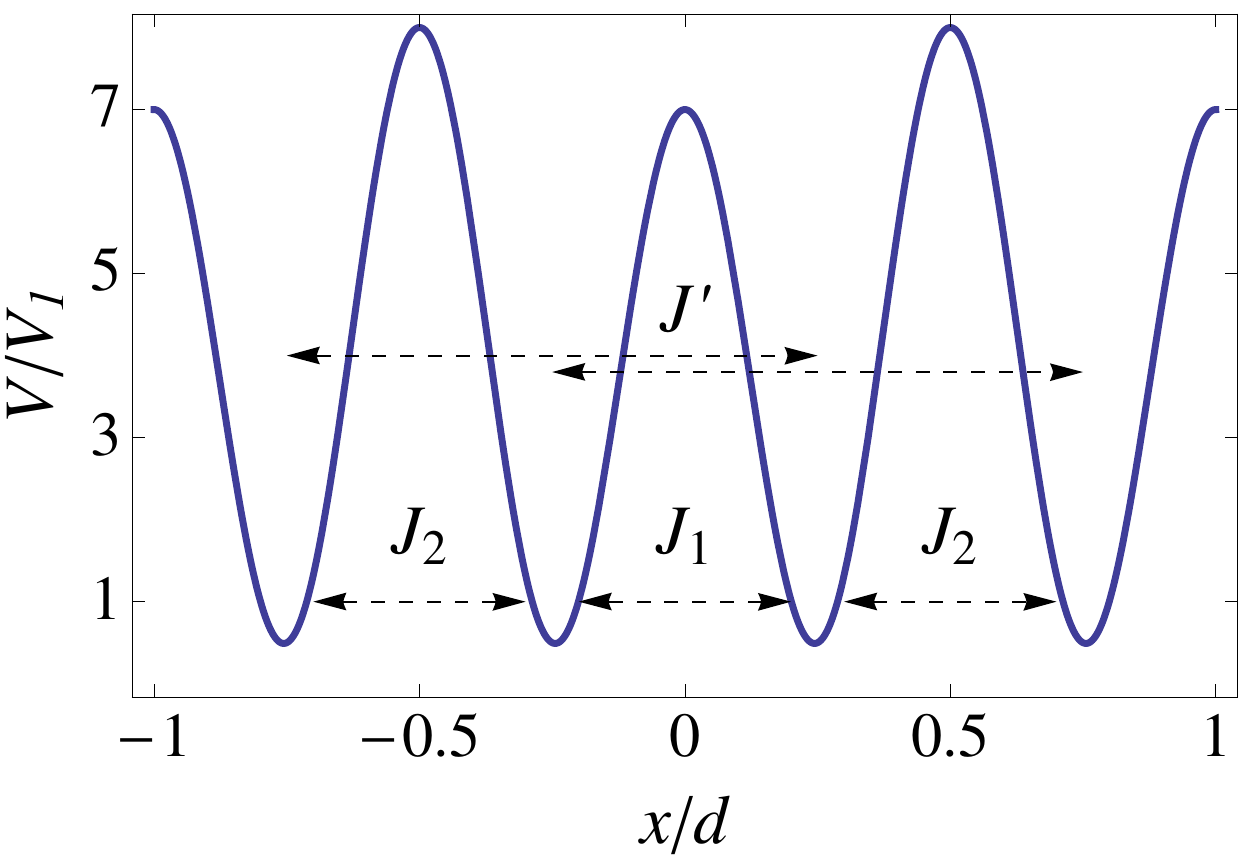}
\begin{flushright}
\includegraphics[width=0.43\textwidth]{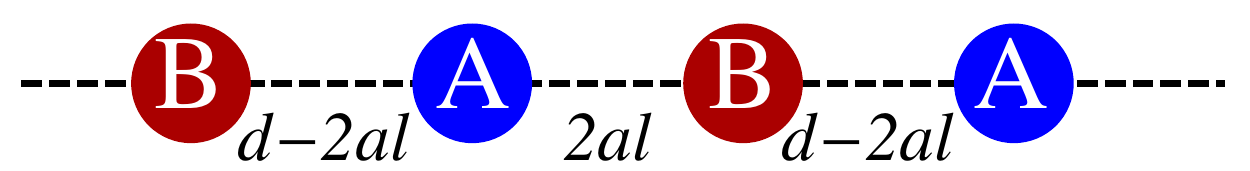}
\end{flushright}
\caption{(Color online) Potential profile for $ V_1=1 E_{\rm{rec}}$ and $V_2=7E_{\rm{rec}} $. The main hopping coefficients used in the tight-binding model are also displayed (see text). For this potential profile, one finds $l=0.2443$.}
\label{potential}
\end{figure}

\begin{figure}[!htbp]
\centering
\includegraphics[width=0.45\textwidth]{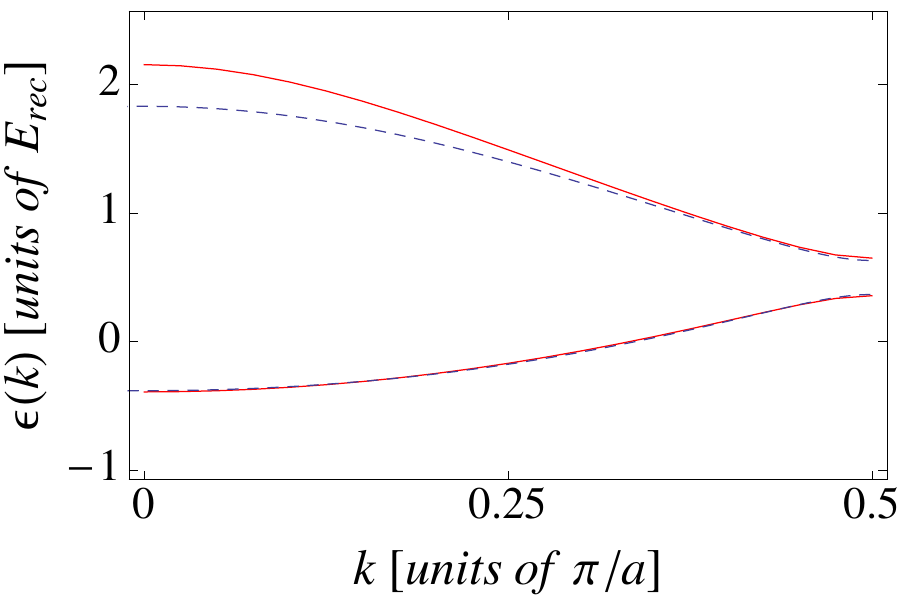}
\caption{(Color online) Lowest two bands for $ V_1=1 E_{\rm{rec}}$ and $V_2=7E_{\rm{rec}} $. The red solid line is the result from the numerical solution of the Schr\"{o}dinger equation; the blue dashed line is the tight-binding spectrum, where the parameters have been chosen by fitting the lowest band.}
\label{band}
\end{figure}

We consider a one-dimensional optical potential of the form \cite{modugno}
\begin{equation}
\label{optpot}
V(x)=V_{1}\sin^2(qx)+V_{2}\sin^2(2qx+\pi/2) ,
\end{equation}
where \(V_1, V_2>0 \) and  \(q= \pi / d \), so that the periodicity of the lattice is \(d\). In Fig.~\ref{potential} we show the shape of such a potential for the choice of parameters $ V_1=1 E_{\rm{rec}}$ and $V_2=7E_{\rm{rec}}$, where $E_{\rm{rec}}=\hbar^2 \pi^2 / 2Md^2$ denotes the recoil energy of atoms with mass $M$. The choice of the phase $\pi/2$ in the optical potential ensures that the bottom of all the wells is at the same depth, while the maxima alternate in height, thus leading to a bipartite lattice. Therefore, the unit cell of the corresponding optical lattice contains two sites, that we denote by $A$ and $B$.  We introduce here a notation that will become useful later: since the spacing between neighboring wells is not constant, but is alternating in length, we call the shortest distance $2al$ and the largest distance $d-2al$, where $a\equiv d/2$ is now the average distance between two neighboring wells.

The aim of this work is to study the optical potential (\ref{optpot}) subject to an external driving that periodically shifts the full potential according to
\be
x\rightarrow x+x_{0} \cos(\omega \tau),
\ee
with $x_0$ the maximum displacement and $\omega$ the frequency of the shaking. Recently, such a problem has been studied for a single atom loaded in the lattice, focusing in particular on the phenomenon of dynamical localization and its consequences on the superfluid-Mott insulator transition for an interacting gas of bosons \cite{hai2014}. In our work, we will instead discuss the effect of the driving term on a system of fermions, for which the presence of a Fermi surface has dramatic consequences already at the non-interacting level. This time-periodic shift of the potential can be realized, for instance, by frequency modulation of the laser beams creating the optical potential \cite{lignier2007}.

We now focus our attention on the potential (\ref{optpot}) in the absence of driving, and let the study of the time-dependent problem to the second part of the present work. To calculate the band structure, it is useful to rewrite the potential as
\be \label{potexp}
V(x)= - \frac {V_1} {4} \left(e^{i2qx} + e^{-i2qx} \right) + \frac {V_2} {4} \left(e^{i4qx} + e^{-i4qx} \right),
\ee
where we have dropped an overall constant. The Schr\"{o}dinger equation for an atom in a space-periodic potential reads
\be
\label{schr}
\left[- \frac {\hbar^2}{2M} \frac {\partial^2} {\partial x^2} + V(x) \right] \psi_{nk} (x) = \epsilon_n(k) \psi_{nk}(x),
\ee
with $\psi_{nk}(x)= e^{ikx}u_{nk}(x)$, where \(n\) is the band index, \( k \) is quasi-momentum, and \( u_{nk}\) are Bloch functions. Since the Bloch functions are periodic with the periodicity $d$ of the lattice, we can perform a Fourier expansion and finally express the wave function as
\be
\label{blochpsi}
\psi_{kn}(x) = \sum_m c_m^{(n)} e^{i(k + \frac {2 \pi} {d} m) x},
\ee
where \(m\in \mathbb{Z}\). By substituting Eq.~(\ref{potexp}) and Eq.~(\ref{blochpsi}) in Eq.~(\ref{schr}), one can cast the Schr\"{o}dinger equation into the form
\bea
\lefteqn{ 4(k + m) ^2 c_m^{(n)} +
\left[- \frac {V_1} {4} \left( c_{m-1}^{(n)} + c_{m+1}^{(n)} \right) \right. } \nn\\
&&+ \left. \frac {V_2} {4} \left( c_{m-2}^{(n)} + c_{m+2}^{(n)} \right) \right] = \epsilon_n(k) c_m^{(n)},
\eea
where we renamed \(ka/ \pi\ra k\), so that \( -1/2 \le k \le 1/2\) and $V_1,V_2$, and $\epsilon_n$ are now expressed in units of $E_{\rm{rec}}$. This equation defines a linear system for the unknown coefficients $c_m^{(n)}$ that can be easily solved with standard libraries.

We have truncated the Fourier expansion retaining $m$ from $-5$ to $5$, corresponding to 11 bands. The result for $ V_1=1 E_{\rm{rec}}$ and $V_2=7E_{\rm{rec}} $ is shown in Fig.~\ref{band}.


\section{Tight-binding model}

The single-particle Hamiltonian in second quantization reads
\be
\hat{H}_0 = \int dx~ \hat{\psi}^\dag(x) \left[ - \fr {\hbar^2} {2M} \fr {\partial^2} {\partial x^2} + V(x) \right] \hat{\psi}(x)\,.
\ee
In this work, we restrict ourselves to a zero-temperature analysis and thus we retain the lowest two bands only, which is a reasonable assumption for sufficiently deep optical lattices, i.e. when $\max\{V_1,V_2\}\gtrsim 5\, E_\rm{rec}$, and when the interactions are weak compared to the energy separation between these two bands and higher ones. One can introduce a set of maximally-localized Wannier functions \cite{ashcroft} centered around the minimum of each well which form a complete single-particle orthonormal basis (further details of how to construct these single-particle states for a bipartite lattice are given in Refs.~\cite{modugno,marzari}). Thus, we can expand the field operators (retaining only the lowest bands states) as
\be
\hat{\psi}(x) = \sum_{j \nu} \hat{a}_{j \nu} \mathcal{W}_0(x-R_{j \nu}),
\ee
where \( \hat{a}_{j \nu}^{(\dag)} \)  destroys (creates) an atom in the Wannier state \( \mathcal{W}_0(x-R_{j \nu}) \) localized at the minimum \( \nu=A,B\) in the  cell \( j \). From now on, we will suppress the double-index notation to identify the lattice sites, in favor of a single-index notation and use the convention that $A$ sites are mapped to even sites. The single-particle tight-binding Hamiltonian is therefore
\bea
\label{tbham}
\lefteqn{\hat{H}_0 = -J_1\sum_{j}(\hat{a}_{2j}^\dagger\hat{a}^{}_{2j+1} +\rm{h.c.} ) }\\
&&-J_2\sum_{j}(\hat{a}_{2j}^\dagger\hat{a}^{}_{2j-1} +\rm{h.c.} ) - J' \sum_{j}(\hat{a}_{j}^\dagger\hat{a}^{}_{j+2} +\rm{h.c.} )\nn
\eea
The definition of the parameters of the model is given in Appendix A. We dropped a term $\sum_j E_j \hat n_j$ because it only leads to an energy shift, given that one can assume the on-site energies in each well to be equal, i.e. $E_A=E_B$ (the wells have the same depth and the same curvature). However, the on-site energy has been determined when fitting the bands (see Table~\ref{hoppcoeff}). Moreover, because of the symmetries of the potential, we assumed the next-nearest-neighbor hopping $A\ra A$ to be equal to $B\ra B$ and we called it $J'$.  The Hamiltonian can be diagonalized in momentum space, yielding the spectrum (in units where we take the lattice spacing $a=1$)
\be \label{etbbb}
\epsilon_{\pm}(k)=-2 J' \cos 2k \pm \sqrt{\Delta(k)}\,,
\ee
where
\be
\label{gap}
\Delta(k) = J_1^2+J_2^2 + 2J_1 J_2 \cos 2k \,.
\ee
We see that the spectrum is invariant under the following two transformations:
\be
J_1 \rightarrow -J_1, \quad J_2 \rightarrow -J_2\,,
\ee
and
\be
J_1 \rightarrow J_2, \qquad J_2 \rightarrow J_1.
\ee
Moreover, one notices that the gap at $k=\pi/2$ is directly connected to the fact that $J_1\neq J_2$. Indeed, were this not the case, i.e. $J_1=J_2$, one would recover the monopartite limit and the gap would close.

\begin{table}[!t]
\centering
\begin{tabular}{| l | l |}
    \hline
    Parameter & Fit \\
    \hline\hline
    \( E_{A,B} \) & 0.612  \\ \hline
    \( J_1 \) & 0.6195  \\ \hline
    \( J_2 \) & 0.4870  \\ \hline
    \( J' \) & -0.0564 \\
    \hline
\end{tabular}
    \caption{Fitting parameters of the tight-binding model for $V_1 =1\, E_{\rm{rec}}$ and $V_2 =7\, E_{\rm{rec}}$. All the parameters are given in units of recoil energy $E_{\rm{rec}}$.}
    \label{hoppcoeff}
\end{table}

The hopping coefficients of the tight-binding model have been estimated by fitting the lowest branch of the spectrum $\epsilon_-(k)$ in Eq.~(\ref{etbbb}) to the numerical results from the band structure calculation. The results for the case $ V_1=1 E_{\rm{rec}}$ and $V_2=7E_{\rm{rec}} $ are summarized in Table \ref{hoppcoeff} and the comparison with the exact band structure calculation is shown in Fig.~\ref{band}. A more accurate estimate of these parameters would require the calculation of the Wannier functions or the use of the method described in Ref.~\cite{modugno}, which is beyond the scope of this work.

\section{Floquet theory}

\begin{figure}[!htbp]
\centering
\includegraphics[width=0.45\textwidth]{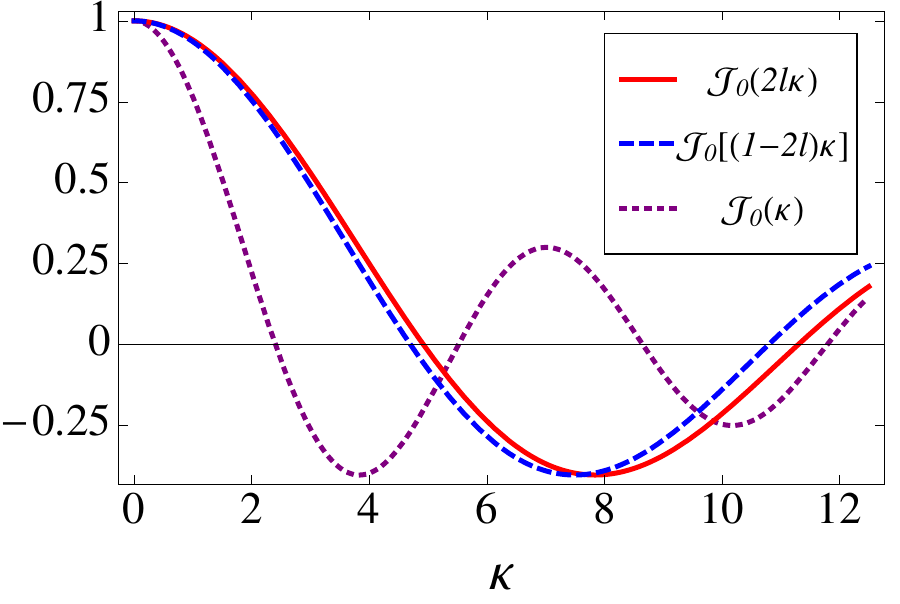}
\caption{(Color online) Relevant Bessel functions renormalizing the hopping coefficients as a function of $\kappa\equiv K/\hbar\om$. As in Fig.~\ref{potential}, we have considered $l=0.2443$.}
\label{bessel}
\end{figure}

Let us now turn to the time-dependent problem and consider a shaken optical potential according to $x\rightarrow x+x_{0} \cos(\omega \tau)$. In the reference frame of the lattice, the single-particle Hamiltonian can thus be written as \cite{eckardt2005,eckardtarimondo}
\be
\hat H(\tau)=\hat H_{0}+\hat W(\tau),
\ee
where the driven part is the dipole term
\be
\hat W(\tau)=\hat xF_{\omega} \cos(\omega \tau), \qquad F_{\omega}=Mx_{0}\omega^2.
\ee
In second quantization, the driven part has the form
\be
\hat{W}(\tau)=F_{\omega} \cos(\omega \tau)\sum_{i,j} \bra R_i | \hat x | R_j \ket \hat{a}_{i}^{\dag} \hat{a}_{j}\,,
\ee
where we defined
\be
\label{dipolewannier}
\bra R_i | \hat x | R_j \ket = \int dx~ \mathcal{W}_{0}^{*}(x-R_i) \,x\, \mathcal{W}_{0}(x-R_j).
\ee
Performing the shift $x \rightarrow x +  (R_{i}+ R_{j} )/2 $ and assuming that the Wannier functions can be chosen real and with a well defined parity (in the present case they can be taken as even functions), one finds that the matrix elements (\ref{dipolewannier}) are vanishing unless $i=j$. Since the Wannier functions are exponentially localized \cite{modugno, marzari}, one obtains
\be
\int dx~ \mathcal{W}_{0}^{*}(x-R_i) \,x\, \mathcal{W}_{0}(x-R_i) \simeq  R_i.
\ee
We choose now the zero of coordinates as in Fig.~\ref{potential} and we thus rewrite the positions of the lattice sites $R_j$ as $R_j= a (j+l_j)$, where \(l_j=-l-j/2\) for $j$ even and \(l_j=l-(j+1)/2\) for $j$ odd.
This leads to a time-dependent term
\be
\hat{W}(\tau)=K \cos(\omega \tau) \sum_{j} (j+l_{j}) \hat{n}_{j \nu}, \qquad K=aMx_{0} \omega^{2}.
\ee
To treat the full time-dependent problem, we use the Floquet theory, valid for Hamiltonians that are periodic in time \cite{sambe,grifoni,hemmerich2010}. We introduce a composite Hilbert space \( \mathcal{H}' = \mathcal{H} \otimes \mathcal{H}_{T} \), where \( \mathcal{H} \) is the original Hilbert space and \( \mathcal{H}_T \) is the Hilbert space of $T$-periodic complex-valued functions. We then define the scalar product in \( \mathcal{H}' \) as
\be
\braket{\braket{\cdot | \cdot}}=\frac{1}{T} \int_{0}^Td\tau\braket{\cdot | \cdot},
\ee
where \(\braket{\cdot | \cdot}\) is the scalar product in \(\mathcal{H}\). According to Floquet theorem, the solutions of the Schr\"{o}dinger equation have the form $|\psi_n(\tau)\ket = e^{-i E_n \tau} |u_n(\tau)\ket $. The quasienergies $E_n$ and the Floquet modes $|u_n(\tau)\ket$ satisfy the eigenvalue problem $\mathfrak{\hat H}(\tau)|u_n(\tau)\ket = E_n |u_n(\tau)\ket$, where $\mathfrak{\hat H}(\tau)\equiv \hat{H}(\tau)-i\hbar\partial_\tau$ is the so-called Floquet Hamiltonian. Moreover, quasienergies that differ by $m\hbar\om$, with $m\in \mathbb Z$, identify the same solution of the Schr\"odinger equation, leading to a Brillouin zone structure. The next aim is to calculate the eigenvalues of the Floquet Hamiltonian. We choose Fock-like states \( | \{ n_{j} \} \ket  \) as a basis of \( \mathcal{H} \), whereas we consider plane waves as a basis of \( \mathcal{H}_T \). The basis vectors in $\mathcal{H'}$ are therefore defined as
\be
 |\{n_{j}\},m \rangle \rangle  = |\{n_{j} \}\ket \exp(im\omega \tau)\,.
\ee
It is now convenient to perform a unitary transformation \cite{eckardt2005} that changes the basis vectors into
\be
|\{n_{j}\} \ket \exp \left[-i\frac{K}{\hbar \omega}\sin(\omega \tau)\sum_{j}(j+l_{j})n_{j}+im\omega \tau\right]\,,
\ee
which is useful to compute the matrix elements of the Floquet Hamiltonian,
\be
\langle \braket{\{n_{j}'\},m' | \hat{H}_0+\hat{W}(\tau) -i \hbar \partial_\tau |\{n_{j}\},m} \rangle.
\ee
We now focus the attention on the hopping terms, i.e \(\hat H_{0}\). They are all of the form
\begin{align} \label{sand}
\langle \braket{\{n_{j}'\},m' | \hat{a}_{i }^{\dag}   \hat{a}_{i'} |\{n_{j}\},m} \rangle =
\langle \{n_{j}'\}| \hat{a}_{i}^{\dag}   \hat{a}_{i'} |\{n_{j}\} \rangle g(T),
\end{align}
\begin{figure}[!htbp]
\centering
\includegraphics[width=.52\textwidth]{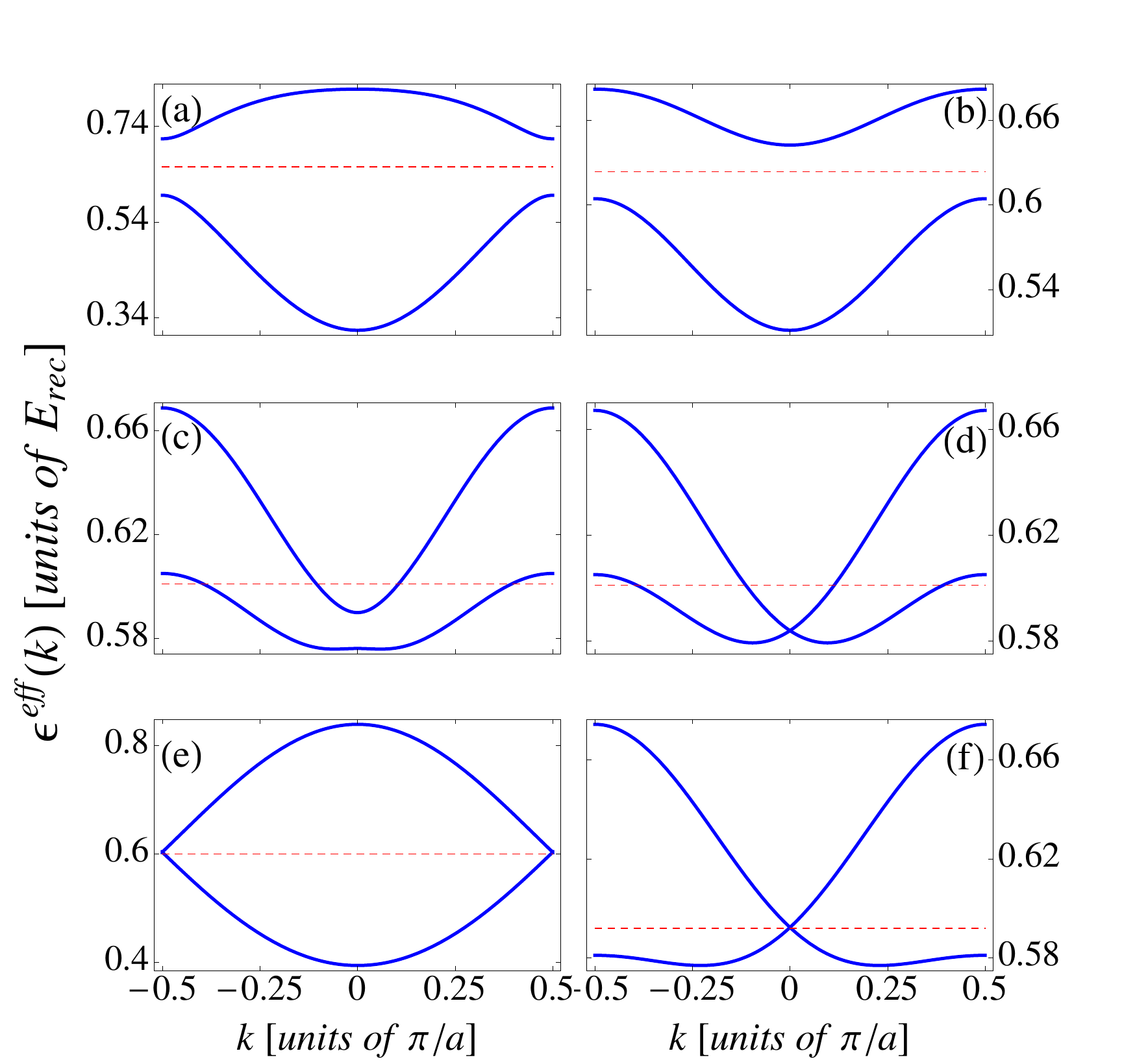}
\caption{ (Color online) Quasienergy spectra of $\hat H_0^{\rm{eff}}$ in units of $E_{\rm{rec}}$ for (a) $K/\hbar\om = 4$, (b) $K/\hbar\om = 4.6$, (c) $K/\hbar\om = 4.8$, (d) $K/\hbar\om = 4.824$, (e) $K/\hbar\om = 5.7$, (f) $K/\hbar\om = 11.074$. The red dashed line is the Fermi level at half-filling.}
\label{bands}
\end{figure}
with
\be
g(T)=\frac{1}{T} \int_{0}^Td\tau \exp \Bigg\{ i\frac{K}{\hbar \omega}s \sin(\omega \tau)
-i(m'-m)\omega \tau \Bigg\}\,, \nonumber
\ee
where we defined $\displaystyle s= \sum_{j}(j+l_{j})(n_{j}'-n_{j})$. By using the integral representation of the Bessel functions of the first kind
\be
\mathcal{J}_{n}(x)=\frac{1}{2\pi} \int_{0} ^{2 \pi} d \tau e^{ix \sin \tau-in \tau},
\ee
we can rewrite Eq.~(\ref{sand}) as
\be
\braket{\{n_{j}'\}| \hat{a}_{i}^{\dag}  \hat{a}_{i' } |\{n_{j}\}}
\mathcal{J}_{m'-m}\left( \frac {K} {\hbar \omega }s \right)\,.
\ee
Only a limited set of matrix elements (\ref{sand}) is needed, because the tight-binding Hamiltonian includes three main hopping processes. Let us consider, as an example, the case $i=2p$ and $i'=2p+1$ with $p$ an integer, which corresponds to the hopping term with amplitude $J_1$. The Fock-state configurations that give non-zero matrix elements are
\bea
\{ n_j \} &=& \{  n_1, n_2, \dots, n_{2p}, n_{2p+1}, \dots \}\,,\\
\{ n_j' \} &=& \{  n_1, n_2, \dots, n_{2p}\pm 1, n_{2p+1}\mp 1, \dots \}\,,
\eea
yielding
\bea
s &=& (2p+l_{2p})(n_{2p}\pm 1 - n_{2p}) \\
&& + (2p+1+l_{2p+1})(n_{2p+1}\mp 1 - n_{2p+1})=\mp 2l\,.\nn
\eea

\begin{figure}[!htbp]
\centering
\includegraphics[width=0.4\textwidth]{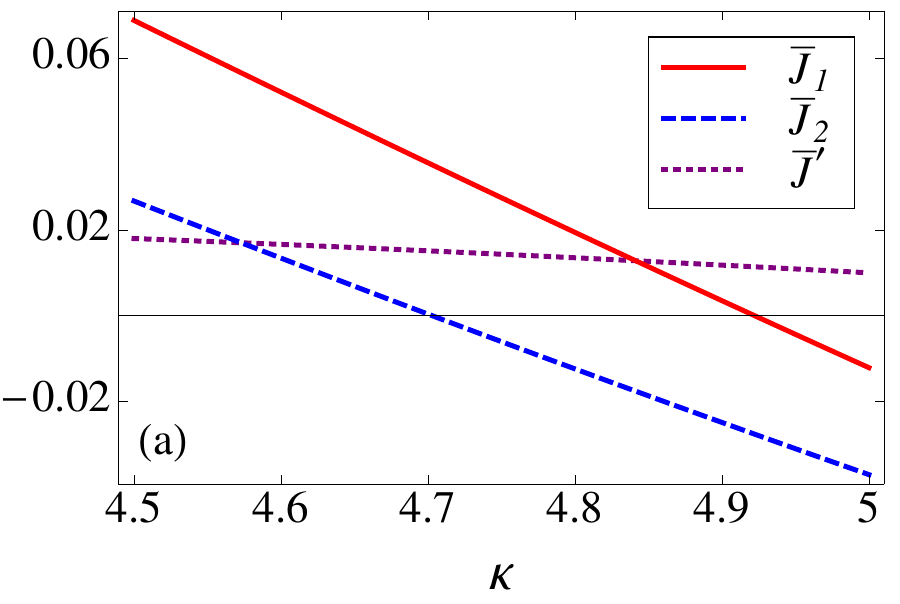}
\includegraphics[width=0.4\textwidth]{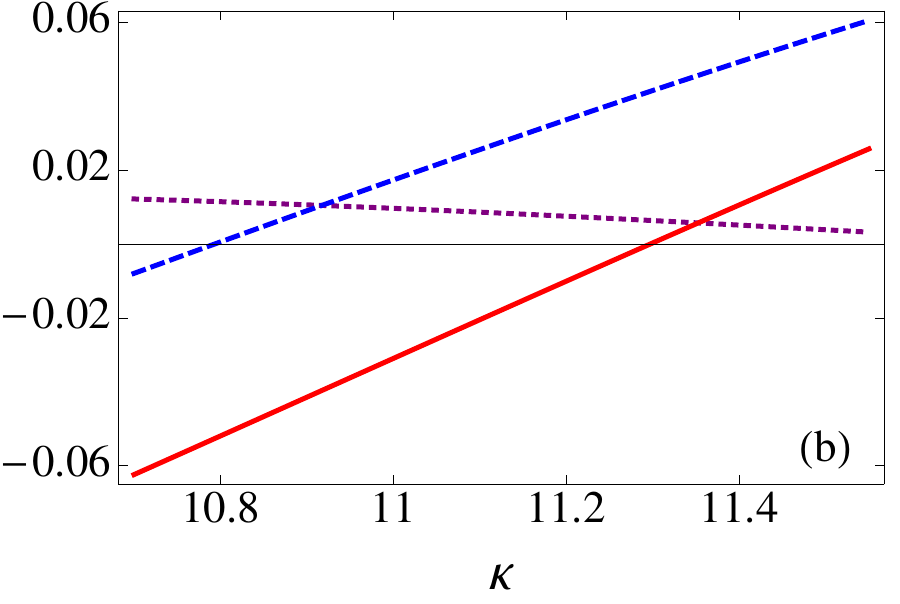}
\caption{(Color online) Renormalized hopping coefficients near the (a) first and (b) second zero of the Bessel functions $\mathcal{J}_0\left( 2l \kappa\right)$ and $\mathcal{J}_0\left[ (1-2l)\kappa\right]$, respectively.}
\label{hopp}
\end{figure}
By using the property that Bessel functions of the first kind with odd index are odd and Bessel functions with even index are even, one can finally write
\be
\mathcal{J}_{m'-m}\left( \frac {K} {\hbar \omega }s \right)=(\mp1)^{m'-m} \mathcal{J}_{m'-m}\left(2l \frac {K} {\hbar \omega }\right).
\ee
Similar arguments can be applied when $i=2p-1$, $i'=2p$ (hopping term $J_2$) and when $|i-i'|=2$ (hopping terms $J'$), leading respectively to $s=\mp (1-2l)$ and $s=\mp 1$. For the matrix elements of the density operator, namely $i=i'$, one finds that $s=0$ and thus $g(T)=\delta_{m,m'}$.

The term $\hat{W}(\tau)$ in the matrix elements now drops because the time derivative term $-i \hbar \partial_\tau$ cancels it. In the limit $\hbar \om \gg J_1,\, J_2,\, J'$ one can perturbatively neglect the off-diagonal elements of the Floquet Hamiltonian with $m\neq m'$ and therefore write the matrix elements in block-diagonal form
\be
\label{effham}
\langle \braket{\{n_{j}'\},m' | \mathfrak{\hat H} |\{n_{j}\},m} \rangle \approx \delta_{m,m'} \bra \{n_{j}'\} | \hat{H}^{\rm{eff}}_0 + m \hbar \om |\{n_{j}\} \ket\,,
\ee
where the operator $\hat{H}^{\rm{eff}}_0$ has the same functional form as $\hat{H}_0$ [see Eq.~(\ref{tbham})], but with renormalized hopping coefficients
\bea
J_1 \quad &\ra & \quad \bar J_1 \equiv \mathcal{J}_0\left( 2l \frac {K} {\hbar \omega}\right) J_1 \,,\\
J_2 \quad & \ra & \quad \bar J_2 \equiv \mathcal{J}_0\left[ (1-2l) \frac {K} {\hbar \omega}\right] J_2 \,, \\
J' \quad  & \ra & \quad  \bar J' \equiv \mathcal{J}_0\left( \frac {K} {\hbar \omega}\right) J' .
\eea
The behavior of the Bessel functions is shown in Fig.~\ref{bessel}. From now on we take $m=0$ in Eq.~(\ref{effham}), thus choosing one specific Brillouin zone for the quasienergies.


\section{Spectrum of the effective Hamiltonian}

The dependence of the renormalized hopping coefficients on the driving parameter $\kappa\equiv K/\hbar\om$ allows for the realization of several regimes, due to fundamental changes in the shape of the quasienergy spectrum of the effective Hamiltonian $\hat H_0^{\rm{eff}}$.
Since for each hopping coefficient the regimes where the Bessel function changes sign occur for different values of the argument $\kappa$, a very rich behavior is expected, with various realizations of band structure configurations. 
Let us consider the different scenarios and discuss the changes of the spectrum as a function of $\kappa$. We concentrate on the half-filled case and investigate the influence of the spectrum on the transport properties of the different ground states realized.

For relatively small values of $\kappa$, the nearest-neighbor hopping coefficients $J_1$ and $J_2$ simultaneously reduce in magnitude, but the shape of the bands is not much affected, as long as these coefficients are large compared with $\bar J'$ (see Fig.~\ref{bands}(a)). Around $\kappa\approx 4.5$, the second band is inverted and the system displays an indirect gap, as shown in Fig.~\ref{bands}(b): the minimum at $k=0$ of the second band is larger in energy than the maximum at $k=\pi/2$ of the lowest band. In Fig.~\ref{bands}(c) we show the case where the minimum at $k=0$ of the second band lowers in energy and the indirect gap now vanishes. This scenario makes possible the realization of a four Fermi-point metallic state.  In Fig.~\ref{bands}(d), the limiting case where the two bands touch at $k=0$ is shown. This requires, from Eq.~(\ref{gap}), that $\Delta(0) = (\bar J_1+ \bar J_2)^2 = 0$, i.e. $\bar J_1= - \bar J_2$, as can be observed by a simple inspection of Fig.~\ref{hopp}(a). This scenario is only realizable because the Bessel functions that renormalize the nearest-neighbor hopping coefficients have different arguments, so that $\bar J_2$ can change sign before $\bar J_1$ does. Since the renormalized hopping coefficients $\bar J_1$ and $\bar J_2$ change with different slope as functions of $\kappa$, they can  therefore become equal, despite the fact that their bare value was different in the undriven case. This happens at $\kappa \approx 5.7$, and causes the closing of the gap at $k=\pi/2$ since $\Delta(\pi/2) = (\bar J_1- \bar J_2)^2$ [see Fig.~\ref{bands}(e)]. Another case appears for larger values of $\kappa$. For $\kappa=11.074$, near the points where $\mathcal{J}_0\left( 2l \kappa \right)$ and $\mathcal{J}_0\left[ (1-2l) \kappa \right]$ vanish, one finds once again that $\bar J_1= - \bar J_2$. The band touching point at $k=0$ is shown in Fig.~\ref{bands}(f).


\section{Phase diagram at half-filling}

\begin{figure}[!htbp]
\centering
\includegraphics[width=0.4\textwidth]{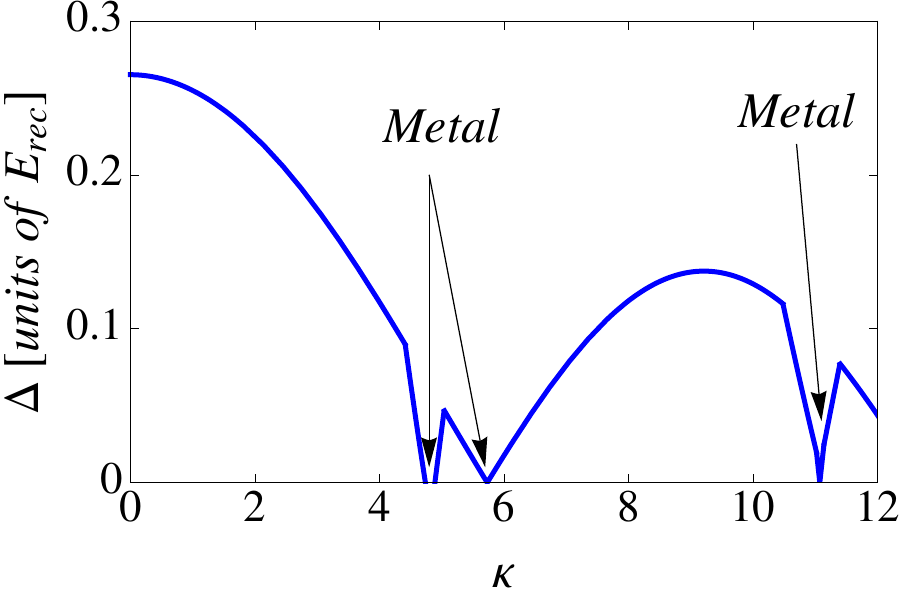}
\caption{ (Color online) Band gap as a function of the shaking parameter $\kappa$ showing several metal-insulator transitions.}
\label{gapPlot}
\end{figure}

By using the band analysis presented in the previous paragraph, we can now describe the behavior of the system in the half-filled case (one particle per site and total (pseudo)spin $S^z_\rm{tot}=0$) in the absence of interactions. In Fig.~\ref{gapPlot} we show how the band gap $\Delta$ changes as a function of $\kappa = K/\hbar\om$.
In the regimes where $\Delta\neq 0$ (which include the undriven case with $\kappa=0$), the Fermi energy lies inside the gap and the system is a Peierls insulator.

One notices that near $\kappa=4.5$, the gap function is not smooth and starts dropping rapidly to zero. The reason for this non-smooth behavior is the inversion of the second band, leading to a change of the gap from direct to indirect. These features in the gap behavior appear for many values of $\kappa$, and are always related to band inversion (either the first or the second band).

The system undergoes two metal-insulator transitions around  $\kappa\approx 4.8$. One can easily prove that the metal phase appears for $\bar J' \geq  \left(  | \bar J_1 + \bar J_2 | - |\bar J_1 - \bar J_2 |  \right)/4$. For the parameters chosen here, this yields $4.74<\kappa<4.89$. In this metal phase, the Fermi surface exhibits four Fermi points as shown in Fig.~\ref{bands}(c)-(d). 

For $\kappa>4.89$, a gap opens again and leads to a (Peierls) insulating behavior. Eventually, the gap at $k=\pi/2$ vanishes at $\kappa=5.7$ where $\bar J_1 = \bar J_2$ and one finds again a metal (see Fig.~\ref{bands}(e)). Moreover, since the nearest-neighbor hopping coefficients are now equal, the unit cell consists of only one lattice site and the Brillouin zone is doubled. Therefore, the spectrum of Fig.~\ref{bands}(e) corresponds to a folded cosine-like band and the metallic phase in this case is the standard two Fermi-point gapless phase of a 1D half-filled electron system in absence of lattice dimerization.

A remarkable behavior seems to occur at $\kappa=11.074$. At this point, the bands approach each other linearly at $k=0$ because the gap closes since $\bar J_1 = -\bar J_2$ (see Fig.~\ref{bands}(f)). However, this dispersion that apparently exhibits one single Fermi point does not lead to a new metallic phase but to a conventional Luttinger liquid with two Fermi points. One can easily reach this conclusion by performing a canonical transformation on the fermionic operators $\hat a_n \ra e^{i \alpha_n} \hat a_n $, where $\{\alpha_n\} \equiv \{ \dots, \pi, \pi, 0, 0,\pi, \pi, 0, 0, \dots \}$. Such a transformation flips the sign of the hopping coefficient every second bond and therefore maps the model with alternating hopping to the typical model with uniform hopping and a cosine-like band, thus leading to a conventional metal with two Fermi points. The price to pay is that the next-nearest neighbor hopping coefficient will also change sign, but since it is quite small in magnitude compared to the nearest-neighbor one, it will have no consequences on the metallic properties.


\section{Effect of interactions}

Let us now add to the Hamiltonian (\ref{tbham}) a Hubbard interaction term
\be
\hat{H}_U = U \sum_i \hat n_{i \ua} \hat n_{i\da}\,,
\ee
that is commonly realized in experimental setups at low temperatures \cite{esslinger}. The Hubbard parameter $U$, defined in  Appendix A, depends on the $s$-wave scattering length $a_s$, and can therefore be tuned by using Feshbach resonances (for example for $^{40}K$ atoms), thus spanning the repulsive regime $U>0$, the attractive regime $U<0$ and the non-interacting limit $U=0$. Since this term has a density-density form, it is not affected by the shaking scheme previously discussed and therefore appears also in the effective Hamiltonian $\hat{H}^{\rm{eff}}_0$, under the supplementary condition that $\hbar \omega \gg U$.

We will focus in the rest of this section on the fate of the region where the unconventional metal with four-Fermi points is found, once the Hubbard interaction is turned on. As we will show, a central role is played by the next-nearest neighbor hopping $\bar J'$. On the other hand, interactions will not affect the positions where the two Fermi-point metals are realized because this involves only a relation between $\bar J_1$ and $\bar J_2$, namely $\bar J_1= \pm \bar J_2$. 

To study the effect of interactions, we first write the non-interacting part in the Peierls form
\bea 
\label{PHam}
\hat{H}_0&=&-t \sum_{n} (1+(-1)^n \delta) \left( \hat{a}_n^{\dagger}\hat{a}_{n+1} + \rm{h. c.} \right)  \nn\\
&& +t' \sum_{n}  \left( \hat{a}_n^{\dagger}\hat{a}_{n+2} + \rm{h.c.} \right)+ \mu \hat N  \nn\\
&=&  \hat H_t +\hat H_{t\delta} + \hat H_{t'} + \mu N\,,
\eea
where we have defined 
\be
t (1+\delta) = \bar J_1\,,\quad t(1-\delta)=\bar J_2\,,\quad t'=-\bar J'\,,
\ee
and a chemical potential has been introduced to control the filling. In the case discussed in this work, the indirect gap is due to the band inversion of the upper band, given by the condition
\be
\bar J' > \bar J'_{c1} = \frac 1 4 \left(  | \bar J_1 + \bar J_2 | - |\bar J_1 - \bar J_2 |  \right) = \frac 1 2 \left( |t| - |t\delta| \right). 
\ee
The transition in the single particle spectrum from a Peierls insulator to a metal with four Fermi points appears when the indirect gap closes, \emph{i.e.}
\be
\bar J'>\bar J'_{c2}=\frac 1 4 \left( | \bar J_1 + \bar J_2 | + |\bar J_1 - \bar J_2 | \right) = \frac 1 2 \left( |t| + |t\delta| \right)\,.
\ee
These critical values are renamed for $t'$ as $t'_{c1}=-\bar J_{c1}'$ and $t'_{c2} = -\bar J_{c2}'$. The transition therefore occurs for $t'<t'_{c2}$. The chemical potential in the Peierls insulator at half-filling is chosen to lie in the center of the (direct or indirect) gap. This defines the chemical potential
\bea
\mu = \left\{ 
\begin{array}{l l }
 |t \delta| - |t| & \quad  t_{c2}' <t' < t_{c1}'\,, \\
2t' & \quad t'>t_{c1}'\,.
\end{array}
\right.
\eea


\subsection{Bosonization}

The theory is bosonized in a similar way as in Ref.~\cite{Japaridze_MH}. One considers the terms $\hat H_{t'}$, $\hat H_{t\delta}$ and $\hat H_U$ as perturbations. The ``unperturbed" spectrum given by $\hat H_t$ is linearized around the Fermi points, that are given at half-filling by $k_F=\pm \pi/2a$;  the corresponding Fermi velocity is $v_F = \pm 2t a$.

In the continuum limit, one substitutes
\be
\hat a_{n,\sigma} \rightarrow \sqrt{a} \, e^{i k_F x} \hat \psi_{R\sigma} (x) + \sqrt{a} \,e^{-ik_F x} \hat \psi_{L\sigma}(x),
\ee
where $x=na$ and $\psi_{L\sigma}(x)$, $\psi_{R\sigma}(x)$ denote, respectively, left and right movers. The fermionic fields are then bosonized according to
\be
\hat \psi_{R(L)\sigma}(x) = \frac{1}{\sqrt{2\pi a}} e^{\pm i \sqrt{\pi}[\hat \phi_\sigma(x)\pm \hat \theta_\sigma(x)]}\,.
\ee
A change of basis for the bosonic fields $\phi_{\sigma}(x)$ and $\theta_{\sigma}(x)$ (from now on we drop the hat on the operators) is performed to describe the charge and spin degrees of freedom:
\bea
\phi_c = \frac{1}{\sqrt{2}}  ( \phi_{\uparrow} +  \phi_{\downarrow})\,,
\quad
\phi_s = \frac{1}{\sqrt{2}}  ( \phi_{\uparrow} -  \phi_{\downarrow}) \\
\theta_c = \frac{1}{\sqrt{2}}  ( \theta_{\uparrow} +  \theta_{\downarrow})\,,
\quad
\theta_s = \frac{1}{\sqrt{2}}  ( \theta_{\uparrow} -  \theta_{\downarrow})\,.
\eea
The Hamiltonian can thus be cast into the following form
\be
H = H_c + H_s + H_{cs}\,,
\ee
where
\bea
H_c &=& \int dx \left\{\enspace  \frac{u_c}{2 K_c}  [\partial_x \phi_{c} (x)]^2  +  \frac{u_c K_c }{2} [\partial_x \theta_{c} (x)]^2 \right.\nn\\  
&& \left. - \mu_{\rm{eff}} \sqrt{\frac{2 }{\pi}} \partial_x \phi_c (x) - \frac{U}{2 \pi^2 a^2}   \cos [\sqrt{8 \pi } \phi_c (x)] \right\}\,, \nn \\
H_s &=&  \int dx \left\{ \enspace  \frac{u_s}{2 K_s}  [\partial_x \phi_{s} (x)]^2  + \frac{u_s K_s}{2}
 [\partial_x \theta_{s} (x)]^2 \right. \nn\\
 &&\left.+\frac{U}{2 \pi^2 a^2}  \cos [\sqrt{8 \pi } \phi_s (x)]\right\} \,, \nn\\
 H_{cs}&=&  -\frac{ 4 t \delta  }{ \pi a}  \int dx \,  \cos[\sqrt{2 \pi} \phi_c(x)]  \cos[\sqrt{2 \pi} \phi_s(x)]\,,
\eea
and we defined $\mu_{\rm{eff}}\equiv \mu - 2t'$, $u_c K_{c}=u_s K_{s}=v_F$, $u_c/K_c=1+U/\pi v_F$ and $u_s/K_s=1-U/\pi v_F$. The bosonic model just derived couples charge and spin degrees of freedom because of the term $H_{cs}$. For this reason, the exact solution of this model is not known and one has to resort to approximation methods or numerical calculations \cite{Japaridze_MH}.


\subsection{Phase diagram analysis}

In the non-interacting limit $U=0$ studied in the previous sections, the half-filled system shows a transition from a band insulator to a metal with four Fermi points. Such a transition, that happens when $t'<t'_{c2}$, can be also predicted in the bosonized model written in terms of $\phi_\sigma$ and $\theta_\sigma$. The condition is that the effective chemical potential $\mu_{\rm{eff}}$ exceeds the mass gap $2t\delta$, which in turn yields $t'<t'_{c2}$ \cite{Japaridze_MH}. In the charge and spin representation, the model becomes rather more complicated but one can obtain a qualitative understanding (also of the interacting case) by performing a mean-field decoupling of the $H_{cs}$ term, along the same lines as in Ref.~\cite{Japaridze_MH}. 

One introduces the expectations values
\bea
\label{mfeq}
m_c &=& 4t\delta \mv{ \cos[\sqrt{2 \pi} \phi_s(x)]}\,, \\
m_s &=& 4t\delta \mv{ \cos[\sqrt{2 \pi} \phi_c(x)]}\,,
\eea
and writes $H = \tilde{H}_c + \tilde{H}_s$ with
\bea
\label{decham1}
\tilde H_c &=& H_c -\frac{ m_c }{ \pi a}  \int dx \,  \cos[\sqrt{2 \pi} \phi_c(x)] \,, \\
\label{decham2}
\tilde H_s &=& H_c -\frac{ m_s }{ \pi a}  \int dx \,  \cos[\sqrt{2 \pi} \phi_s(x)]  \,,
\eea
which now displays a clear decoupling between charge and spin degrees of freedom. However, the new mass terms still couple the two sectors thanks to the mean-field equations (\ref{mfeq}). There is an asymmetry in the charge sector due to the presence of the effective chemical potential $\mu_{\rm{eff}}$, which is responsible for the phase transition from metal to insulator as previously argued for the non-interacting case. In the weak coupling limit $U\ll t$, where $K_{c,s}\approx 1$, the terms proportional to $\cos [\sqrt{8 \pi } \phi_{c,s} (x)]$ can be neglected because they are irrelevant and the new terms $\cos [\sqrt{2 \pi } \phi_{c,s} (x)]$ dominate the physics of this system. One can therefore analyze the model in the form (\ref{decham1})-(\ref{decham2}) by using the exact solution found by Zamolodchikov \cite{zamo} for the sine-Gordon Hamiltonian with $\beta^2=2\pi$
\bea
\lefteqn{H^{SG}_\alpha =  }\\
 && \int dx \left\{  \frac{u_\alpha}{2}  [ (\partial_x \phi_\alpha)^2  +
 (\partial_x \theta_\alpha)^2]- \frac{m_\alpha}{\pi a}  \cos [\sqrt{2 \pi K_\alpha } \phi_\alpha]\right\} \,,\nn
\eea
when $0<K_\alpha<2$ and $\alpha=c,s$. Here the Luttinger parameter $K_\alpha$ has been reabsorbed into newly defined bosonic fields $\phi_\alpha \ra \sqrt{K_\alpha}\, \phi_\alpha$ and $\theta_\alpha \ra \theta_\alpha / \sqrt{K_\alpha}$. The excitation spectrum consists of solitons, antisolitons and breathers (soliton-antisoliton bound states). The lowest-energy excitations in this range of $K_\alpha$ are given by the breathers. The lightest breather mass $\Delta_\alpha$ (which is twice the energy gap of the system) is related to the soliton mass $M_\alpha$ via
\be
\Delta_\alpha = 2M_\alpha\sin\left(\frac{\pi}{2} \frac{K_\alpha}{4-K_\alpha}\right)\,.
\ee
The soliton mass $M_\alpha$ can be calculated from the bare mass $m_\alpha$ using the relation
\be
M_\alpha/\Lambda = C(K_\alpha) \left(m_\alpha/\Lambda\right)^{2/(4-K_\alpha)}\,,
\ee
where $\Lambda$ is a high-energy cut-off. Finally, to solve the mean-field equations one needs \cite{lukyanov}
\be
\langle \cos(\sqrt{2\pi K_\alpha}\phi_\alpha)\rangle = B(K_\alpha)(M_\alpha/\Lambda)^{K_\alpha/2}\,.
\ee
The coefficients $C(K_\alpha)$ and $B(K_\alpha)$ are given in Appendix B. Based on this approach, one can solve the self-consistent equations for the charge and spin gaps, $\Delta_c$ and $\Delta_s$ respectively, and obtain a qualitative understanding of the role of interactions. For $\mu_{\rm{eff}}=0$ the two gaps are equal when $U=0$. The charge gap increases as a function of $U$, while the spin gap decreases. Therefore, repulsive interactions lead to a larger charge gap, while they reduce the spin gap. This picture is confirmed by numerical simulations \cite{kampf}, but a quantitative agreement would require a careful estimate of the Luttinger parameters, which is beyond the scope of the present work.

Let us now consider the effect of the chemical potential $\mu_{\rm{eff}}$ on the four Fermi-point phase. Such a phase appears for 
$\kappa_A < \kappa < \kappa_B$, where $\kappa_A \approx 4.74$ and $\kappa_B \approx 4.89$. In the non-interacting picture, the transition occurs when $\mu_{\rm{eff}}$ exceeds the band gap. One can assume an analogous criterion to hold in the interacting case, \emph{i.e.} $\mu_{\rm{eff}}>\Delta_c/2$, where $\Delta_c$ is the lowest breather mass in the charge sector, as discussed above. In the presence of the interactions the charge gap is renormalized and increases as a function of $U$, as concluded already at the mean-field level. Therefore, the critical value of $t'$ for the metal transition changes because the effective gap to overcome now depends on $U$, and for repulsive interactions it is larger than for $U=0$. One thus expects that the interval $[\kappa_A, \kappa_B]$ shrinks because the charge gap that $t'$ needs to overcome has now increased. In the limit of strong Hubbard coupling ($U\gg t,t'$) the charge 
gap $\Delta_{c} \sim U$ and the range of $\kappa$ where the metallic phase is reached vanishes above a critical value $U_c$, i.e. when the charge gap is large enough, such that the effect of $t'$ is no longer sufficient to close it. On the other hand, attractive interactions $U<0$ have the opposite effect. In the limit of strong Hubbard coupling ($|U|\gg t,t'$) the charge gap $\Delta_{c} \sim \delta t^{2}/|U|$ and therefore the region  where the metallic phase is realized enlarges. 

As it follows from the performed mean-field analysis, in the case of weak repulsive interaction  and in close proximity to the metal-insulator transition ($t^{\prime} \leq t^{\prime}_{c2}$), the charge gapless phase is also spin gapless and thus shows properties of a Luttinger liquid. However, deeply inside the metallic phase ($t^{\prime} \ll t^{\prime}_{c2}$), where the properties of the system are determined by the four Fermi points and the effect of the direct single-particle gap is negligible, the system becomes similar to the one-dimensional half-filled $t-t'$ Hubbard model. This model has been studied in detail, both analytically and numerically \cite{Fabrizio_96,Kuroki_97,Fabrizio_98,DaulNoack_98,DaulNoack_00, AebBaerNoack_01, Gros_01,Gros_02,Torio_03,Gros_04, Fabrizio_04}, and it is known to give rise to a Luther-Emery liquid for attractive and repulsive on-site interactions, i.e. a spin gapped metal. 

In the repulsive case, the dominant instability is the charge-density-wave, which exhibits the slowest power-law decay of the corresponding correlations. Notice that this behavior is different from the conventional Hubbard model, for which the charge gap is open, the spin gap is zero and the dominant correlation is the spin-density-wave. In the opposite case of attractive on-site coupling, the spin gap is present for arbitrary $t^{\prime} < t^{\prime}_{c2}$ and the system behaves as a spin gapped metal with dominant singlet-superconducting instability, characterized by a power-law decay of the corresponding correlations.


\section{Conclusions}

In this paper, we investigate how to realize metal-insulator transitions for a system of fermionic atoms loaded in a bipartite one-dimensional optical lattice at half-filling. The bipartite character of the optical lattice is essential because it ensures that the nearest-neighbor hopping coefficients alternate in magnitude, opening a gap at the edge of the Brillouin zone ($k=\pi/2$). The Fermi level lies inside the gap at half-filling and therefore the system behaves as a band insulator (Peierls insulator).

By introducing an external high-frequency driving force that shakes the lattice, we show that the hopping coefficients are all renormalized by Bessel functions that depend on the shaking parameter $\kappa$ with different arguments. This feature allows for a competition of the different hopping coefficients, which can reduce in magnitude and change sign, severely altering the shape of the bands. We observe that the system can exhibit band inversion, generating an indirect gap, as well as band touching and band crossing.

The different regimes reached by this scheme show several possible transitions from Peierls insulators with direct or indirect gap to metallic states with two or four Fermi points. The scheme discussed in this work represents, to the best of our knowledge, the first method that has been proposed to experimentally realize such an unconventional four Fermi-point metallic state, the properties of which have been theoretically discussed in the literature in the past decades \cite{Fabrizio_96,Kuroki_97,Fabrizio_98,DaulNoack_98,DaulNoack_00, AebBaerNoack_01, Gros_01,Gros_02,Torio_03,Gros_04, Fabrizio_04}. Notice that this cannot be realized in conventional lattices, the bipartite nature of the lattice being an essential requirement.

Finally, we qualitatively investigate the effect of on-site interactions on the metallic phases. The two Fermi-point metallic phase, appearing only at some discrete values of the driving parameter $\kappa$, behaves as an ordinary Luttinger liquid and therefore is expected to be analogous to the conventional Hubbard model.  Concerning the four Fermi-point metallic phase, we argue, based on a mean-field analysis supported by former numerical calculations, that the region in $\kappa$ where such a phase appears would shrink (and eventually disappear) for repulsive interactions, whereas it would widen for attractive ones. A quantitative estimate of this process is left for future investigations.


\section*{Acknowledgements} We would like to thank M. \"Olschl\"ager, A. Hemmerich, T. Comparin and D. Gerace for useful discussions. This work was supported by the Netherlands Organization for Scientific Research (NWO) and is part of the D-ITP consortium, a program of the NWO that is funded by the Dutch Ministry of Education, Culture and Science (OCW). GIJ acknowledges the Georgian National Science  Foundation and the Science and Technology Center in Ukraine for support through the project  STCU-5893.

\addcontentsline{toc}{section}{Bibliography}
\bibliographystyle{mprsty}
\bibliography{Biblio}

\appendix

\section{Tight-binding parameters}
Define the single particle Hamiltonian in first quantization as
\be
\hat H_0 = - \frac {\hbar^2}{2M} \frac {\partial^2} {\partial x^2} + V(x)\,.
\ee
The definition of the parameters of the tight-binding Hamiltonian can be written as
\bea
J_1 &=& - \int dx\,\,  \mathcal{W}_0^{*}(x-R_{j A})\, \hat H_0 \, \mathcal{W}_0(x-R_{jB})\, \\
J_2 &=& - \int dx \,\,  \mathcal{W}_0^{*}(x-R_{j A}) \, \hat H_0 \,\mathcal{W}_0(x-R_{(j - 1)B})\, \\
J' &=& - \int dx \,\,  \mathcal{W}_0^{*}(x-R_{j \nu})\, \hat H_0 \, \mathcal{W}_0(x-R_{(j+1) \nu})\,  \\
E_{\nu} &=&  \int dx\,\,  \mathcal{W}_0^{*}(x-R_{j \nu}) \, \hat H_0\, \mathcal{W}_0(x-R_{j \nu})\,.
\eea
In the presence of $s$-wave interactions, the Hubbard parameter $U$ introduced in the main text has the form
\be
\label{Uterm}
U = \frac 1 2 \times \frac{4\pi \hbar^2 a_s^{\rm{eff}}}{M} \int dx\,\, | \mathcal{W}_0(x-R_{j\nu})| ^4\,,
\ee
where $a_s^{\rm{eff}}$ is the effective $s$-wave scattering length for the 1D system, therefore containing also the contribution of the harmonic confinement in the two orthogonal spatial directions.

One can understand the reason why the Hubbard parameter $U$ does not carry a sublattice index by considering the harmonic approximation. Since the two wells have the same curvature, the corresponding harmonic oscillator states (\emph{i.e.} the Wannier functions) have the same form in the two wells and the integral in Eq.~(\ref{Uterm}) will be independent on which well is referred to.

On the other side, the hopping parameters are determined by an overlap integral, and since the wells have a relative distance that alternates in magnitude, the hopping will differ and will alternate in magnitude accordingly.

\section{Parameters for the sine-Gordon exact solution}

The exact solution of the sine-Gordon Hamiltonian given in the main text contains the two parameters $C(K_\alpha)$ and $B(K_\alpha)$ that are given by 
\bea \label{C-K} C(K_\alpha)&=&
\frac{2\Gamma(\frac{K_\alpha}{8-2K_\alpha})}{\sqrt{\pi}\Gamma(\frac{2}{4-K_\alpha})}\cdot
\left[\frac{\Gamma(1-K_\alpha/4)}{2\Gamma(K_\alpha/4)}\right]^{\frac{2}{4-K_\alpha}}
\eea
and
\bea 
\lefteqn{B(K_\alpha) = [\Gamma(1/2+\xi/2)\Gamma(1-\xi/2)]^{(K_\alpha/2)-2}\times} \nn\\
&& \times\left[\frac{2\sin(\pi\xi/2)}{4
\sqrt{\pi}}\right]^{K_\alpha/2}\left[\frac{(1+\xi)\pi^{2} \Gamma(1-K_\alpha/4)}{
\sin(\pi \xi) \Gamma(K_\alpha/4)}\right] 
\eea
where  $\xi=K_\alpha/(4-K_\alpha)$ and $\Gamma(x)$ is Euler's gamma function.

\end{document}